\documentclass[%
 reprint,
superscriptaddress,
showkeys,
 amsmath,amssymb,
 aps,
pre,
]{revtex4-1}
\usepackage{enumerate}
\usepackage{graphicx}
\usepackage{dcolumn}
\usepackage{bm}
\usepackage{bbold} 
\usepackage{graphicx}
\usepackage{dcolumn}
\usepackage[left=3cm,top=2.5cm,right=3cm,bottom=2.5cm]{geometry}
\usepackage{array}
\usepackage{epsfig}
\usepackage{wrapfig}
\usepackage{xcolor}
\usepackage{soul}
\usepackage{braket}
\usepackage{verbatim}
\usepackage{mathtools}
\usepackage{mathrsfs}
\usepackage{amsfonts}
\usepackage[english]{babel}
\usepackage{textcomp}
\usepackage{float}
\usepackage{comment}
\excludecomment{toexclude}
\usepackage{changepage} 

\begin{document}


\title{Pressure and flow statistics of Darcy flow from simulated annealing}


\author{Marise J. E. Westbroek}
\affiliation{Department of Earth Science and Engineering, Imperial College London, London SW7 2BP, United Kingdom}
\affiliation{The Blackett Laboratory, Imperial College London, London SW7 2AZ, United Kingdom}

\author{Peter R. King}
\affiliation{Department of Earth Science and Engineering, Imperial College London, London SW7 2BP, United Kingdom}

\author{Dimitri D. Vvedensky}
\affiliation{The Blackett Laboratory, Imperial College London, London SW7 2AZ, United Kingdom}

\author{Ronnie L. Schwede}
\affiliation{Shell Global Solutions International B.V., Grasweg 31, 1031 HW Amsterdam, The Netherlands}






\begin{abstract}
The pressure and flow statistics of Darcy flow through a random permeable medium are expressed in a form suitable for evaluation by the method of simulated annealing.  There are several attractive aspects to using simulated annealing:~(i) any probability distribution can be used for the permeability, (ii) there is no need to invert the transmissibility matrix which, while not a factor for single-phase flow, offers distinct advantages for the case of multiphase flow, and (iii) the action used for simulated annealing is eminently suitable for coarse graining by integrating over the short-wavelength degrees of freedom.  In this paper, we show that the pressure and flow statistics obtained by simulated annealing are in excellent agreement with the more conventional finite-volume calculations.



\end{abstract}

 
                            
\maketitle


\section{Introduction}
\label{Intro}

The study of flow in porous media has a wide range of applications, including hydrology (e.g.~\cite{bear, mualem}), geothermal engineering (e.g.\cite{abdelazizac}), materials science (e.g.~\cite{bear}), and the medical sciences (e.g.~\cite{khaled}). Another application, and the focus of the work reported here, is the flow of oil in a reservoir.

There are two basic ways to conceptualize the flow through a porous material. One approach is to solve the Navier-Stokes equations.  On domains in the millimeter to centimeter scale, the fluid configurations can be imaged with X-ray microcomputed tomography \cite{liu17}. The flow of oil through a rock, by contrast, calls for flow descriptions on the kilometer scale.  Solving the Navier--Stokes equations on the later scale is not feasible because of limited information about the rock permeability and the matrix form in which to cast the problem. But, even if all of this information was available, the computational requirements would be prohibitive.

On a more coarse grained scale, such as the slow flow of a viscous fluid, the Navier-Stokes equations can be reduced to Darcy's law, a relation between the effective permeability $K$, the average velocity (``flow'') $\bf{q}$ of the fluid, and the pressure gradient $\nabla \bf{p}$:
\begin{equation}\label{Darcy}
{\bf q}=-K \nabla {\bf p}\, .
\end{equation}
Here, $K({\bf x})=k({\bf x})/{\mu}$, where $k({\bf x})$ is the effective permeability of the medium and $\mu$ is the viscosity of the fluid. 
The effective permeability describes the medium on a ``mesoscopic'' scale, large compared to the pore scale, but small on the scale of the macroscopic medium. Although proposed as an empirical relation by Darcy in the 1850s \cite{darcy}, Darcy's law can be recovered from the Navier-Stokes equations \cite{whitaker}. In this paper, we focus on single-phase, incompressible flow: 
\begin{equation}\label{incompressibility}
\nabla \cdot \bf{q}=0\, .
\end{equation}
In reservoir engineering, Darcy's law is solved numerically using the finite-volume method \cite{aziz, schafer}, while the finite-element method is common in hydrology.

In previous work \cite{westbroek}, we formulated the solution to one-dimensional Darcy flow as a path integral over pressure. In discrete form, the path integral is a tool to simulate Darcy pressure paths $\{p_i\}$ on a spatial lattice using Markov chain Monte Carlo methods according to the weighting $e^{-S}$, where the ``action'' $S=S[\{p_i\}]$ contains the solution to Darcy's law.

In one-dimension, the path integral hinges on an analytical solution obtained by combining (\ref{Darcy}) and (\ref{incompressibility}),
\begin{equation}
{d\over dx}\biggl(K{dp\over dx}\biggr)=0\, ,
\label{eq3}
\end{equation}
from which we immediately obtain
\begin{equation}
K{dp\over dx}=q_0\, ,
\label{eq4}
\end{equation}
where $q_0$ is a constant.  In higher dimensions, the equation corresponding to (\ref{eq3}),
\begin{equation}
\nabla\cdot(K\nabla p)=0\, ,
\label{eq5}
\end{equation}
has the general solution
\begin{equation}
K\nabla p=\nabla\times{\bf f}+\nabla g\, ,
\end{equation}
where ${\bf f}$ is a twice differentiable vector function and $\nabla^2g=0$. This solution is not as amenable to the path integral formulation as (\ref{eq4}), so we turn to an alternative expression for the action that utilizes simulated annealing \cite{kirkpatrick,press}.

Although solving the problem using simulated annealing is more computationally intensive than the finite-volume method, there are several advantages over various other techniques.  Simulated annealing allows any probability distribution to be used for the permeability, as does the finite-volume method. However, there is no need to invert the transmissibility matrix during simulated annealing, which, while not an issue single-phase flow, is advantageous for multiphase flow.  On a more formal level, the action used for simulated annealing is eminently suitable for integration over the short-range degrees of freedom to derive coarse-grained permeability coefficients \cite{hristopulos,attinger,eberhard,hanasoge,king2}.  The porous medium can also be characterized as a statistically homogeneous continuum with local fluctuations in physical parameters.  The resulting path integral expression can be averaged over parameter fluctuations to obtain large-distance parameters that describe the flow \cite{tanksley}.

Our paper is organized as follows. Path integral formulations of Darcy flow through random porous media are derived in Sec.~\ref{secII}.  We explain why the procedure we have developed previous for the numerical evaluation of path integrals in one dimension cannot be extended to higher dimensions. Section~\ref{secIII} discusses computation methods, including the simulation of the permeability fields and the method of simulated annealing. Results are presented in Sec.~\ref{secIV}. We have calculated empirical probability densities for the pressure, Cartesian flow components, the total flow along the $y$-direction, and examined the effect of the variance of the log-permeability.  Conclusions and a discussion are provided in Sec.~\ref{secV}, including an assessment of the viability of simulated annealing and the extension of the path integral approach to multi-phase flow.


\section{Theory}
\label{secII}

\subsection{Effective permeability}

The effective permeability is modelled as a stochastic process. Various such models exist, with an overview is given in \cite{renard}. 
We have opted to model $K({\bf x})$ as a lognormal process,
\begin{equation}\label{permeability}
K({\bf x})=e^{L({\bf x})},
\end{equation}
where $L({\bf x})$ is a Gaussian process with mean zero, characterized by its variance $\sigma^2$ and correlation length $\xi$.  More information on this process can be found in \cite{long1D}. This conventional choice \cite{law} has the advantage of a strictly positive permeability.  We emphasize, however, that our method applies to any choice of permeability distribution.

\subsection{Path integral in higher dimensions} 

In an earlier paper \cite{westbroek}, we developed a path integral for Darcy flow in one dimension, The integral is over all discrete pressure trajectories that (subject to the boundary conditions) follow Darcy's law, which is enforced by a delta-functional: 
\begin{align}\label{DeltaFunctional}
&Z=\int\prod_i dp_i\int\prod_i{d L_i}
\exp\biggl[-\sum_{ij}L_i({\mathsf C}_L^{-1})_{ij}L_j \biggr]\nonumber\\
&\quad\times e^{-\sum_i L_i}
\delta\biggl({p_i-p_{i-1}\over\Delta x}+q_0 e^{-L_i}\biggr)\, .
\end{align}
The term
\begin{equation}
\exp\biggl[-\sum_{ij}L_i({\mathsf C}_L^{-1})_{ij}L_j \biggr]
\end{equation}
encodes the correlated Gaussian probability distribution of the log-permeability, where $C_L$ denotes the log-permeability covariance matrix.
The factor $e^{-\sum_i L_i}$ is the Jacobian associated with integrating over the $L_i$ rather than over the $e^{L_i}$.
Upon integration over the $L_i$ we obtain a path integral 
\begin{equation}
Q(\{p_i\})={e^{-S(\{p_i\})}\over Z}\, ,
\label{eqP9}
\end{equation}
with the discrete ``action'' \cite{westbroek}
\begin{align}
&S(\{p_i\})=\sum_i\log\biggl({p_{i-1}-p_i\over q_0\Delta x}\biggr)\nonumber\\
&+\sum_{ij}\log\biggl({p_{i-1}-p_i\over q_0\Delta x}\biggr)({\mathsf C}_L^{-1})_{ij}\log\biggl({p_{j-1}-p_j\over q_0\Delta x}\biggr)\, .
\label{eqP10}
\end{align}
Discrete pressure paths are generated according to the probability density (\ref{eqP9}).

An analogous path integral to (\ref{DeltaFunctional}) in two dimensions is obtained with the standard procedure for classical statistical dynamics \cite{phythian77,peliti78,jouvet79,jensen81}:
\begin{align}
&Z_{\mathrm{2D}}=\int\prod_{ij} dp_{ij}\int\prod_{kl}{d L_{kl}}
\exp\bigg(-\sum_{ij} L_{ij}\bigg)\nonumber\\
&\quad\times\exp\bigg[-\sum_{ij,kl}L_{ij}({\mathsf C}_L^{-1})_{ij,kl}L_{kl}\biggr]\nonumber\\
&\quad\times
\delta\left\{ 
\frac{\partial}{\partial x}\left( e^{L_{ij}} \frac{\partial p_{ij}}{\partial x}\right)+
\frac{\partial}{\partial y}\left( e^{L_{ij}} \frac{\partial p_{ij}}{\partial y}\right)
\right\}.
\label{eq12}
\end{align}
The delta-function enforces the discrete form of Darcy's law (\ref{eq5}), where we have used the notation
\begin{widetext}
\begin{align}
\frac{\partial}{\partial x}\left( e^{L_{ij}} \frac{\partial p_{ij}}{\partial x}\right)&=
\frac{1}{\Delta x}\left[ e^{L_{i,j}} \left( \frac{p_{i,j}-p_{i-1,j}}{\Delta x}\right)
- e^{L_{i-1,j}}\left(\frac{p_{i-1,j}-p_{i-2,j}}{\Delta x}\right)\right]\nonumber\\
&=\frac{1}{\Delta x^2}\left[e^{L_{i,j}}p_{i,j}-(e^{L_{i,j}}+e^{L_{i-1,j}})p_{i-1,j}+e^{L_{i-1,j}}p_{i-2,j}\right]\nonumber\\
\frac{\partial}{\partial y}\left( e^{L_{ij}} \frac{\partial p_{ij}}{\partial y}\right)&=
\frac{1}{\Delta y}\left[ e^{L_{i,j}} \left( \frac{p_{i,j}-p_{i-1,j}}{\Delta y}\right)
- e^{L_{i-1,j}}\left(\frac{p_{i-1,j}-p_{i-2,j}}{\Delta y}\right)\right]\nonumber\\
&=\frac{1}{\Delta y^2}\left[e^{L_{i,j}}p_{i,j}-(e^{L_{i,j}}+e^{L_{i-1,j}})p_{i-1,j}+e^{L_{i-1,j}}p_{i-2,j}\right]\nonumber\\
\end{align}
\end{widetext}

\vskip-10pt

\noindent
The next step is to represent the delta-function as the limit of an exponential, so that the exponentials in (\ref{eq12}) can be combined into a single exponential whose argument is the ``action''.  The usual procedure \cite{phythian77,peliti78,jouvet79,jensen81} is to apply a functional Fourier transform, which yields a {\it complex} action.  This is appropriate for formal studies involving perturbation expansion, where the complex action yields real results despite the complex nature of intermediate calculations.  But the Markov chain Monte Carlo method relies on real variables from the outset, so we represent the delta functional as the limit of a Gaussian probability density:
\begin{widetext}
\begin{align}\label{DeltaFunctional2DGauss}
Z_{\mathrm{2D}}=\int\prod_{ij} dp_{ij}\int\prod_{kl}{d L_{kl}}
\exp\bigg(&-\sum_{ij} L_{ij}\bigg)\exp\bigg[-\sum_{ij,kl}L_{ij}({\mathsf C}_L^{-1})_{ij,kl}L_{kl}\biggr]\nonumber\\
\quad\times\lim_{t\to0}\bigg\{\exp\biggl[{1\over t}\biggl(
&\frac{e^{L_{i,j}}p_{i,j}-(e^{L_{i,j}}+e^{L_{i-1,j}})p_{i-1,j}+e^{L_{i-1,j}}p_{i-2,j}}{\Delta x^2}\nonumber\\
+
&\frac{e^{L_{i,j}}p_{i,j}-(e^{L_{i,j}}+e^{L_{i-1,j}})p_{i-1,j}+e^{L_{i-1,j}}p_{i-2,j}}{\Delta y^2}
\biggr)\biggr]\bigg\}\, .
\end{align}
\end{widetext}
This expression is readily generalized to three dimensions. 

Averages of pressure and correlation functions can be calculated from (\ref{DeltaFunctional2DGauss}) by first generating permeability fields, then setting $t$ to some value, and finally using the Metropolis--Hasting (MH) algorithm to minimize the discrete ``action'':
\begin{equation}
\sum_{ij}\biggl\{\frac{\partial}{\partial x}\left( e^{L_{ij}} \frac{\partial p_{ij}}{\partial x}\right)+
\frac{\partial}{\partial y}\left( e^{L_{ij}} \frac{\partial p_{ij}}{\partial y}\right)\biggr\}.
\end{equation}
Successively smaller values of $t$ are chosen until there is convergence of the pressure distributions. This procedure, which requires separate calculations for each value of $t$, is not especially efficient.  In the next section, we develop a more elegant approach based on simulated annealing.

\subsection{Path integral for simulated annealing}

To account for any number of dimensions, let us write the action in its continuum form:
\begin{equation}
S={1\over2}\int_V K({\bf x})\left(\nabla p({\bf x})\right)^2dV\, ,
\label{action}
\end{equation}
where the integral is carried out over the entire volume under consideration. Simulated annealing aims to find the pressure $p({\bf x})$ that minimizes the action (\ref{action}) for fixed permeability $K({\bf x})$.

We show that the minimized pressure follows Darcy's law. The objective is to extremize the action (\ref{action}) with respect to the pressure.  We vary the action with respect to $p$ by adding an infinitesimal pressure $\delta p$ and imposing the condition
\begin{equation}
S[p({\bf x})+\delta p({\bf x})]-S[p({\bf x})]=0\, .
\label{stationarity}
\end{equation}
Any boundary conditions are unchanged, so $\delta p({\bf x})=0$ at the boundaries of the volume $V$.  If there exists a $p^\ast$ such that the stationarity condition (\ref{stationarity}) holds, the action $S$ is stationary at $p^\ast$.  Retaining $\delta p$ only to first order and performing an integration by parts, we obtain
\begin{align}
&S[p+\delta p]-S[p]=\int K \left(\nabla p\right) \left(\nabla \delta p\right)\,dV\nonumber \\
&=-\int \nabla \cdot \left(K \nabla p \right)\delta p\,dV-\int_{\partial V}K\delta p\nabla p\cdot d{\bf S}\, .
\end{align}
The boundary term vanishes because of Dirichlet boundary conditions fix the pressure across the entry and exit surfaces, and the absence of pressure fluctuations along surfaces perpendicular to the flow direction (Sec.~\ref{secIII}). Because $K$ is always nonzero, the condition (\ref{stationarity}) translates into 
\begin{equation}
\nabla \cdot (K\nabla p)=0,
\end{equation}
which is Darcy's law (\ref{eq5}) for incompressible flow. Thus, if $p^\ast$ can be found such that the stationarity condition is met at fixed $K$, then $p^\ast$ follows Darcy's law.  The simulated annealing algorithm can be applied to the action (\ref{action}) to solve for the pressure. The method is inspired by the process of annealing, which is a treatment whereby a solid is slowly cooled until its structure is eventually frozen at its minimum free energy configuration \cite{du}.


\section{Computational methods}
\label{secIII}

We will follow the convention for units used by the hydrology community.  Darcy's law (\ref{Darcy}) gives the relation between the effective permeability $K~\mathrm{[L^2\,T^{-1}]}$, the flow $\bf{q}~\mathrm{[L\,T^{-1}]}$ and the pressure (also known as the hydraulic head) $p~\mathrm{[L]}$. The total flow in a given direction is denoted by $Q_i~\mathrm{[L^3\,T^{-1}]}$ $(i=x,y,z)$.  Our three-dimensional, rectangular prismatic porous medium is simulated on a grid of 
\begin{equation}
N_x \times N_y \times N_z = 50 \times 70 \times 50
\end{equation}
lattice elements, representing a domain of size
\begin{equation}
X \times Y \times Z = 40\, \mathrm{m} \times 85\, \mathrm{m} \times 25\, \mathrm{m}.
\end{equation}
The correlation lengths of the permeability field are
\begin{equation}
\lambda_x=8\,\mathrm{m};~\lambda_y=8\,\mathrm{m};~\lambda_z=5\,\mathrm{m}.
\end{equation}
We have used Dirichlet boundary conditions along the $y$-direction, making that the main flow direction. No-flow boundary conditions were imposed along the other boundaries. The values chosen for the log-permeability variance $\sigma^2$, the defining feature of the six parameter sets, are given in Table \ref{table:params}. The values in this table were those used by Nowak {\it et al.}~\cite{nowak_main}, which enables a qualitative comparison to be made between our two approaches.

\begin{table}[b]
\begin{ruledtabular}
\caption{\label{table:params} Values used for the variance $\sigma^2$ of the log-permeability.}
\begin{tabular}{ c d l }
\multicolumn{1}{l}{No.} & \multicolumn{1}{c}{Log-permeability variance $\sigma^2$} & \multicolumn{1}{l}{Color}\\ \hline
1  & \multicolumn{1}{d}{0.125} & \textcolor{gray}{\bf gray}\\ 
2 & \multicolumn{1}{d}{0.25} & \textcolor{red}{\bf red}\\ 
3 & \multicolumn{1}{d}{0.5} & \textcolor{blue}{\bf blue} \\
4 & \multicolumn{1}{d}{1.0} & \textcolor{magenta}{\bf magenta} \\
5 & \multicolumn{1}{d}{1.75} & \textcolor{brown}{\bf brown} \\
6 & \multicolumn{1}{d}{2.5} & \textcolor{orange}{\bf orange}\\
\end{tabular}
\end{ruledtabular}
\end{table}

\subsection{Simulation of the permeability fields}

To simulate the permeability field, the first step is to generate the three-dimensional log-permeability field.  Once the Gaussian field $L({\bf x})$ is available, definition (\ref{permeability}) can be invoked to calculate the permeability field $K({\bf x})$. To generate $L({\bf x})$, we have made use of the circulant embedding technique \cite{gneiting}, where the correlation matrix ${\sf C}$ of the desired field is embedded into a matrix ${\sf M}$ that has a circulant or block circulant structure.  Products of the square root ${\sf M}^{1/2}$ with white noise random vectors are realizations of the desired random field \cite{nowak, dietrich}. This method relies on the fast Fourier transform (FFT).  For a $d$-dimensional rectangular mesh containing $N_x^d$ points, the computational requirements are those of an FFT of a vector of dimension $2 N_x^d$ per realization \cite{dietrich}.

\subsection{Simulated annealing}

The simulated annealing algorithm is based on the MH algorithm, a step-by-step explanation of which can be found in \cite{westbroek2}.  In the present case, simulated annealing seeks to minimize the action (\ref{action}). Clearly, the minimum attainable value is zero. The algorithm consists of the following steps.

\begin{enumerate}

\item Initialize a random pressure that is consistent with the boundary conditions.

\item Execute the MH algorithm some $M \gg 1$ times. The MH algorithm lowers the value of the action $S$, but also accepts some modifications to the pressure that increase the action. It explores the entire ``state space'' (the set of values of $S$ as a function of $p({\bf x})$) and does not get stuck in a local minimum of the state space.

\item\label{step3}
After every $N_s \gg 1$ steps, check the value of $S$. When the value of $S$ starts to fluctuate around a constant value, go to step \ref{step4}.

\item\label{step4}
Adapt the MH accept/reject criterion to ``accept the change in the action with probability $\mathrm{min}(1,e^{-\delta S/T})$ for some constant $0<T<1$''. This is a ``cooling step'' \cite{dimensionAction}. The state space is explored in smaller steps than was the case for the standard MH algorithm, while maintaining a constant acceptance rate. The lower the value of $T$, the smaller the steps. In our context, $T$ does not have the interpretation of a temperature, but its effect remains that of slowing down the state space exploration.

\item Repeat steps \ref{step3}-\ref{step4} until the action attains a critically low value $\epsilon_1$, say, $\epsilon_1=0.1$.

\item Employ a modification of the MH algorithm known as the ``greedy algorithm'', which accepts only changes to the pressure that lower the action, until $S$ dives below a second critical value $\epsilon_2$, say $\epsilon_2=10^{-2}$.

\end{enumerate}

To expedite the simulated annealing algorithm, we made use of a technique known as over-relaxation (OR). The idea behind over-relaxation is to choose trial changes that cause significant changes to the pressure field, but only small changes to the action \cite{creutz2,brown}. Such strategic updates combine a thorough exploration of the phase space with a high probability of acceptance. Because the action (\ref{action}) is quadratic, it is possible to calculate an update that leaves the action unchanged, rendering the Metropolis accept/reject step unnecessary. The update
\begin{equation}
p_i^{\mathrm{new}}=2 p_i^\ast-p_i^{\mathrm{old}}
\end{equation}
lies ``on the other side'' of the minimum of the action:~$p_i^\ast$ is the value of $p_i$ that minimizes $S$, with all other parameters kept fixed.
Since the over-relaxation procedure is deterministic, we alternate between OR and regular MH steps to avoid any dependence of the pressure field on its random starting configuration.  Here, we exchanged three in four Metropolis sweeps for OR sweeps.

We have used an exponential cooling scheme
\begin{equation}
T^{(k)}=\alpha^k T_i\, ,
\end{equation}
where $k$ indicates the cooling step.  The MH algorithm was executed $M=2,000$ times for all parameters. For $\sigma^2 \leq 1$, repeating the cooling algorithm $N_s=3,000$ times was found to be a good choice.  For $\sigma^2 > 1$, it was necessary to set $N_s=6,000$. To be able to calculate empirical probability density functions for the pressure and flow, we have worked with $N=10,000$ realizations for each parameter set.

In comparing the computational efforts involved in running the FVM method and simulated annealing, we note that both require a permeability field as input.  The computational cost associated with the FFT is 
\begin{equation}
\mathcal{O}(2 N_x N_y N_z \log (N_x N_y N_z))
\end{equation}
floating point operations (``flops''). The key calculation in the FVM is a sparse matrix inversion. The sparse matrix solver UMFPACK \cite{UMFPACK} can solve such an equation in 
\begin{equation}
\mathcal{O}(N_x N_y N_z \log (N_x N_y N_z))
\end{equation}
flops.  Contrary to the FVM, the simulated annealing requires $\mathcal{O}((N_x N_y N_z)^2)$ flops to calculate a pressure realization.  One factor $N_x N_y N_z$ arises from the number of lattice sites.  The number of required intermediate updates $N_{\mathrm{sep}}$  introduces a further factor $N_x N_y N_z$.  However, there are techniques whose implementation is likely to decrease the run time considerably, such as the multigrid Monte Carlo (MGMC) method \cite{goodman,janke,liu} and directed sampling \cite{duncan1,duncan2,leli}.


\section{Results}
\label{secIV}

We have calculated empirical probability densities for the pressure $p$, the flow components $q_y$ and $q_x$, and the total flow $Q_y$ and looked at the effect of the variance of the log-permeability $\sigma^2$. Based on information about boundedness, we have made parametric fits using the choices made in \cite{nowak_main} for guidance.

All quantities were normalized for straightforward interpretation. The Dirichlet boundary conditions were chosen to yield a pressure difference 
\begin{equation}
\Delta p=1\,\mathrm{m}\, .
\end{equation}
The flow components were normalized as
\begin{align}
q_{x,y}^\ast&={q_{x,y}\over K_e I_0}\, ,\\
\noalign{\vskip3pt}
Q_y^\ast&= {Q_y\over K_e I_0 A}\, ,
\end{align}
where $K_e$ is the theoretical expectation value of the permeability,  $I_0=\Delta p/L_y$ and $A=L_x \times L_z$. Computationally, the normalization was achieved by setting $K_e=1/(I_0 A)$.

\begin{figure}[t]
\centering
\includegraphics[width=.75\columnwidth]{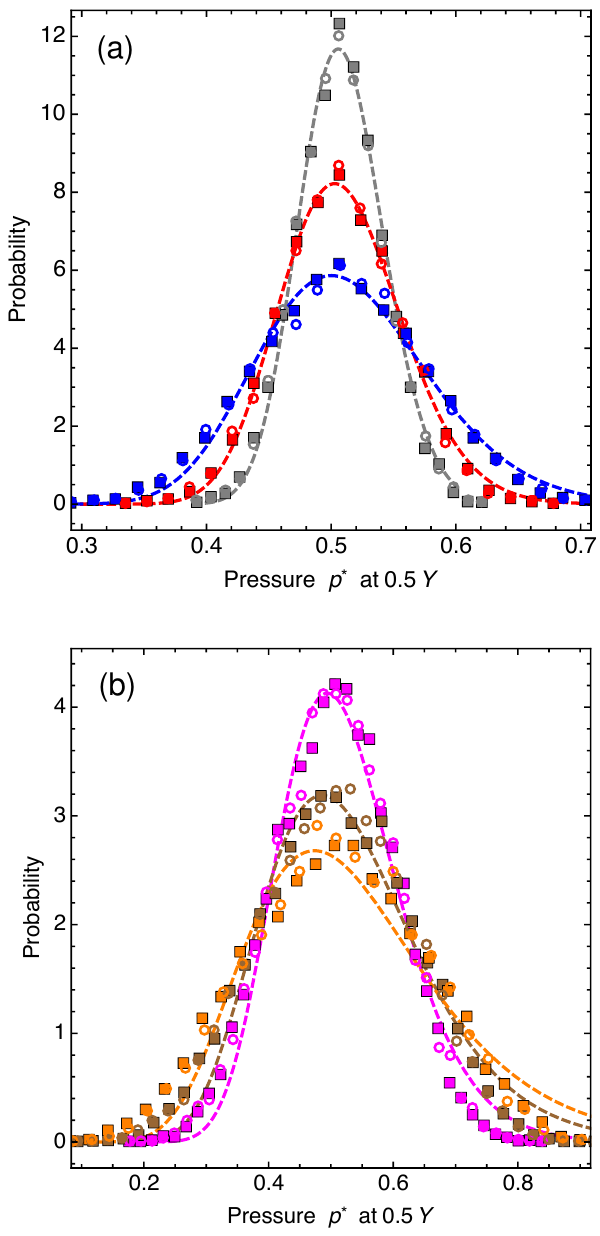}
\caption{Pressure statistics for Darcy flow at the center of the domain, for the following values of the log-permeability variance $\sigma^2$: (a)
$0.125~\mathrm{(grey)}, 0.25~\mathrm{(red)}, 0.5~\mathrm{(blue)}\}$ and (b) $1.0~\mathrm{(pink)}, 1.75~\mathrm{(brown)}, 2.5~\mathrm{(orange)}$, as listed in Table \ref{table:params}. Squares represent the results of the finite-volume method, circles those of simulated annealing.}
\label{fig:hydraulic_head_05}
\end{figure}

\begin{figure}
\centering
\includegraphics[width=.75\columnwidth]{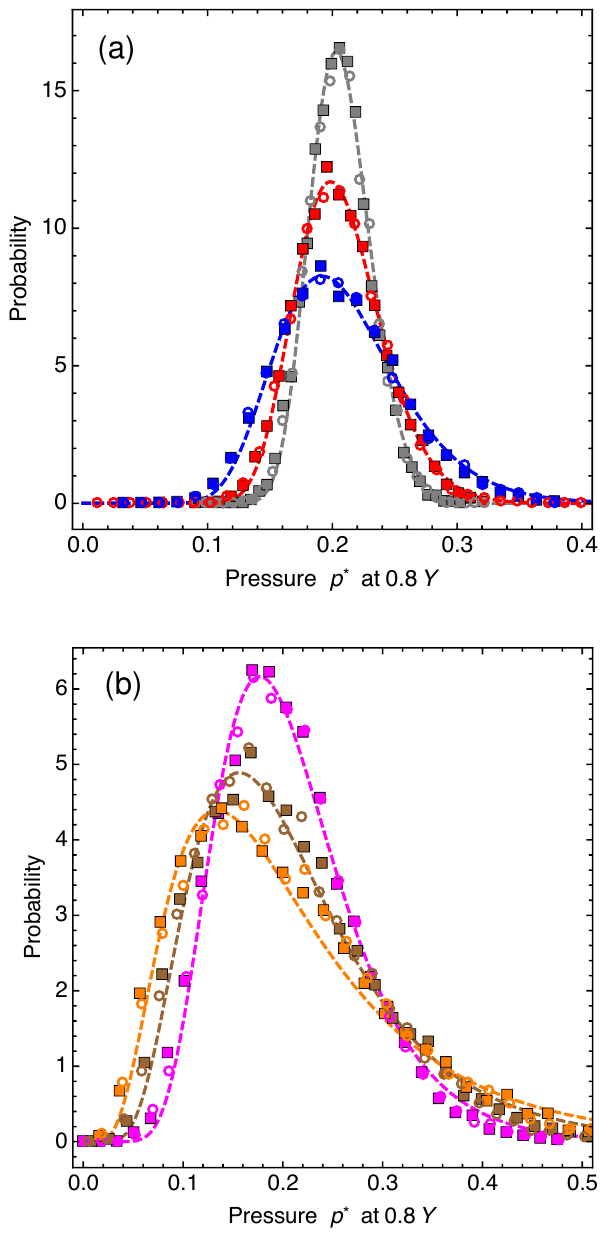}
\caption{Pressure statistics for Darcy flow at the point $(0.5X,0.8Y,0.5Z)$ for finite-volume simulations (squares) and simulated annealing (circles).}
\label{fig:hydraulic_head_08}
\end{figure}

Due to correlations in the permeability, the distributions of pressure and flow are in general not expected to be Gaussian, especially when the variance of the log-permeability is high.  The pressure in the main direction takes values in the interval $[0,1]$, due to the Dirichlet boundary conditions. Given this constraint and the choice of stochastic model for the permeability, the log-normal distribution is an obvious contender for parametric fits to the pressure.
The probability density function is given by:
\begin{equation}
f(x)=\frac{1}{x \sqrt{2\pi} \sigma'}
\exp\left[-\frac{(\log x-\mu')^2}{2(\sigma')^2}\right]\, .
\end{equation}
In order to visualize the dependence of the pressure on its position along the main axis, we have made empirical probability density plots at two positions: $(0.5X,0.5Y,0.5Z)$, which is the center of the domain (Fig.~\ref{fig:hydraulic_head_05}) and $(0.5X,0.8Y,0.5Z)$ (Fig.~\ref{fig:hydraulic_head_08}). From the two sets of figures it is apparent that the lognormal distribution is most evident near the boundary. Towards the center of the domain, the histograms bear more resemblance to the normal distribution \cite{ababou}, as the generalized Central Limit Theorem predicts \cite{bouchaud}.  For a more extensive explanation of this theorem in the context of Darcy flow, see \cite{long1D}. These boundary effects increase with the  log-permeability variance, as can also be observed for the flow.

\begin{figure}
\centering
\includegraphics[width=.75\columnwidth]{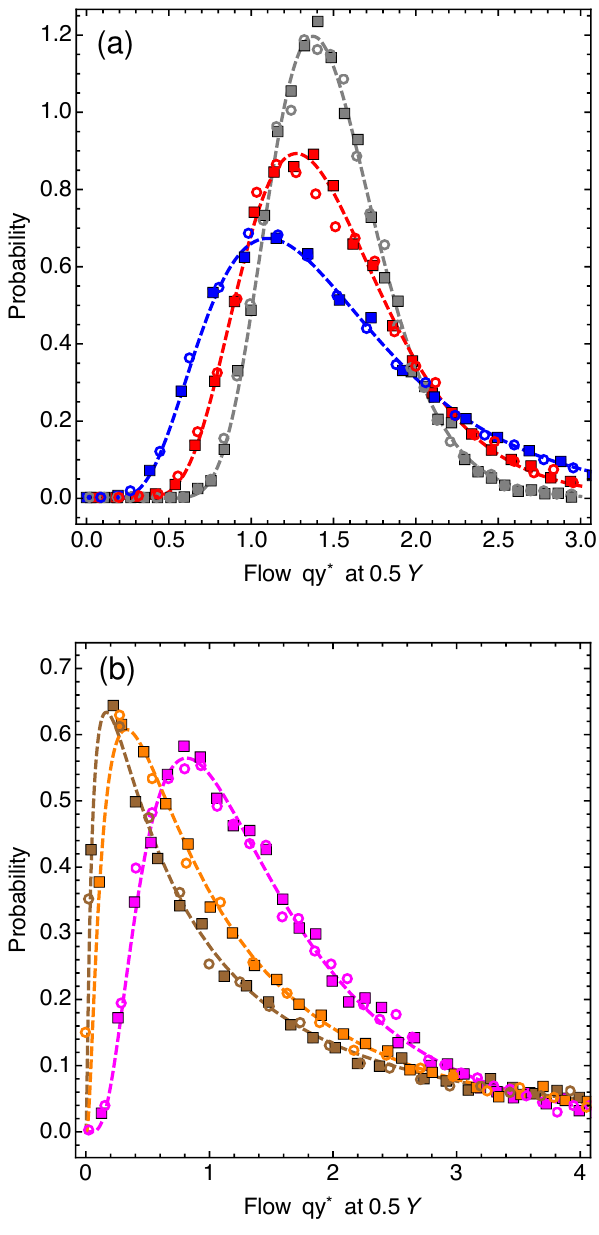}
\caption{Normalized flow in main direction $q_y^*$ for the log-permeability variances $\sigma^2$ given in Table \ref{table:params}, measured at the point $(0.5 X, 0.5Y, 0.5 Z)$. The log-normal pattern manifests itself most clearly for high values of $\sigma^2$. For these high values, the boundary effects, which dictate the log-normality, have the strongest influence on the flow statistics.}
\label{fig:panel_qy_05}
\end{figure}

Like the pressure, the flow along the main axis is subject to a non-negativity constraint, which enforces a one-sided bound. The flow could only be negative in the unlikely event of flow reversal due to locally very high permeability. The observed values for the cases considered in this work were non-negative. For the flow, as for the pressure, the shape of the probability density depends on the vicinity to a restricting boundary. We have evaluated the flow in the main direction at the center of the domain. The results are shown in Fig.~\ref{fig:panel_qy_05}. 

\begin{figure}
\centering
\includegraphics[width=.75\columnwidth]{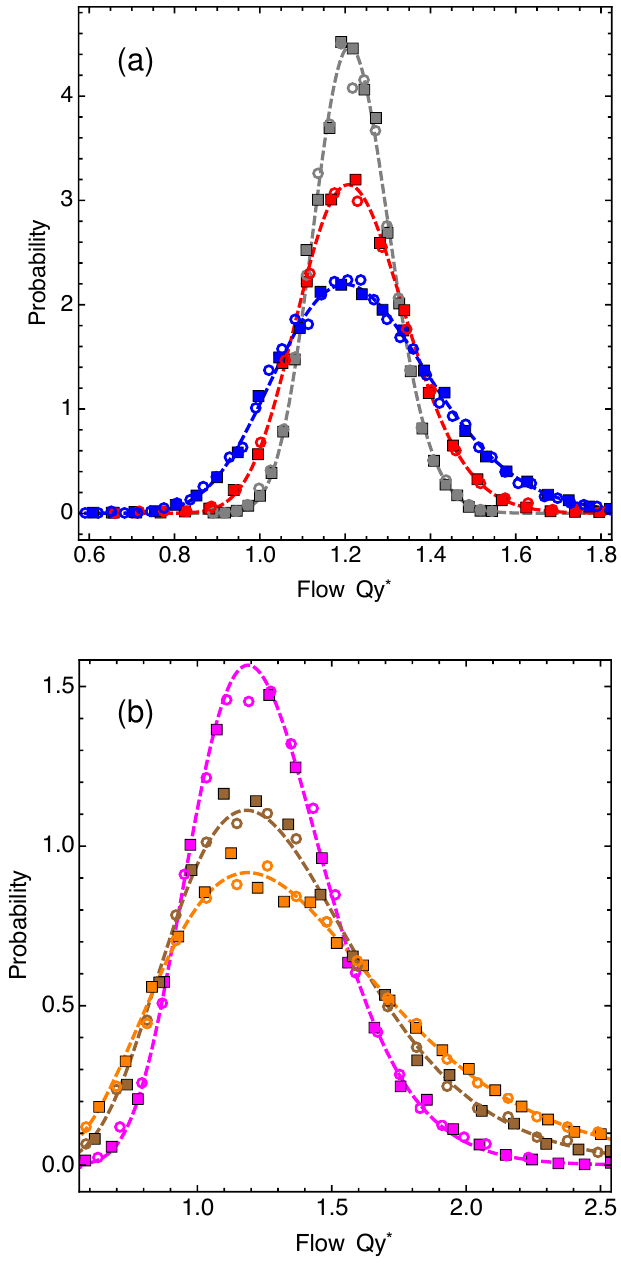}
\caption{Normalized total flow in the main direction $Q_y^\ast$ for the values of $\sigma^2$ stated in Table \ref{table:params}. The distribution of $Q_y^\ast$ resembles that of $q_y^\ast$ shown in Fig.~\ref{fig:panel_qy_05}. Due to the averaging over a cross-section, the distributions appear more Gaussian.}
\label{fig:panel_Qy_08}
\end{figure}

One can see that for high values of the log-permeability variance, the flow statistics are most clearly log-normal.  This is because the boundary effects are more strongly felt for high values of $\sigma^2$, a pattern that can also be observed by comparing Figures \ref{fig:hydraulic_head_05} and \ref{fig:hydraulic_head_08}. The total flow in the main direction, defined as the average over a cross-section perpendicular to the $y$-axis, is conserved along the $y$-axis. The results can be seen in Fig.~\ref{fig:panel_Qy_08}. When comparing Figs.~\ref{fig:panel_qy_05} and \ref{fig:panel_Qy_08}, an obvious difference is that the flow statistics of the total flow approximate the Gaussian distribution more closely. The log-normal distribution tends to the normal distribution in the limit $(\sigma'/\mu')^2 \to 0$. The Gaussian appearance is a result of the averaging over a cross-section that defines the total flow.

\begin{figure}
\centering
\includegraphics[width=.75\columnwidth]{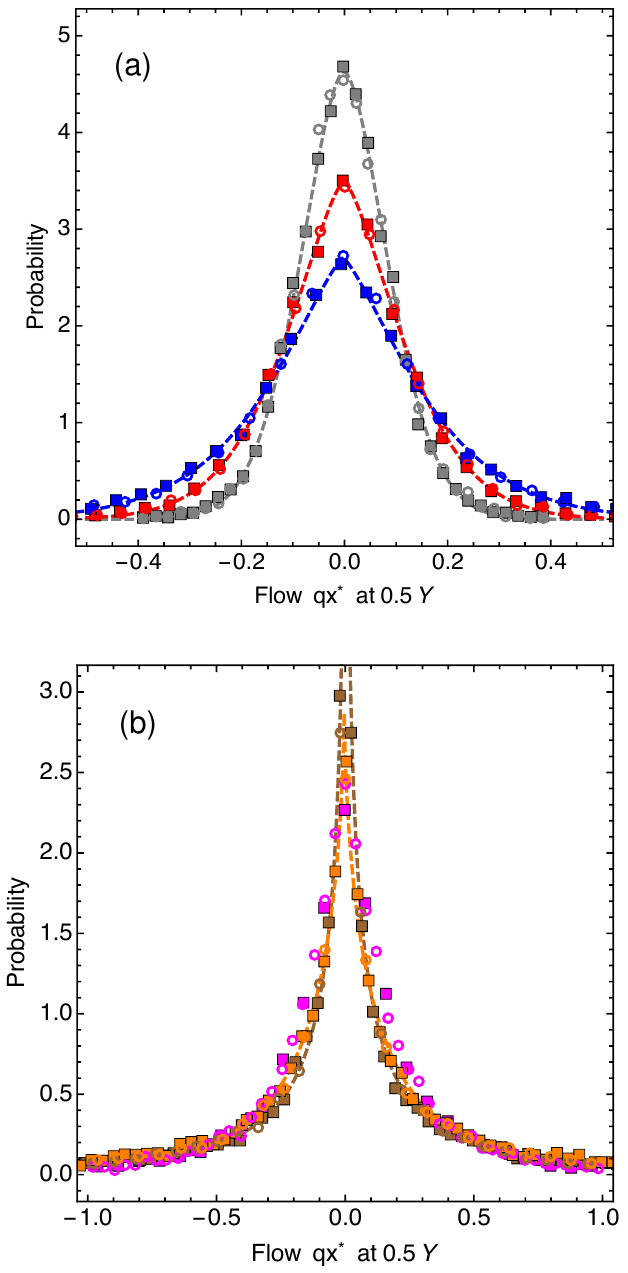}
\caption{Statistics of the normalized flow in the $x$-direction, $q_x^\ast$, again for the parameter set given in Table \ref{table:params}. The parametric fits of the exponential power distribution reflect the symmetry of the statistics about zero. The tails become heavier for greater values of $\sigma^2$. In the limit $\sigma \gg 1$, the flow either continues along the main axis or ``hits a wall'' and reverses course. Thus, in this limit, the likelihood of small values for $q_x^*$ is very small.}
\label{fig:qx}
\end{figure}

For the transverse flow $q_x^*$, shown in Fig.~\ref{fig:qx}, we fitted an exponential power distribution. This choice reflects the expectation that the transverse flow is symmetric about zero.
The probability density function for the exponential power law is:
\begin{equation}
g(x)=\frac{1}{2\sigma''\Gamma(1+1/k)}
\exp\left[-\frac{|x-\mu''|^k}{(\sigma'')^2}\right].
\end{equation}
The Gaussian distribution corresponds to the case $k=2$. All parametric fits for the parameters $\{\mu',\mu'',\sigma',\sigma'',k\}$ were made using the in-built routine {\tt FindFit} of {\sc Mathematica} \cite{mathematica}.

A striking feature of the transverse flow statistics are the long and heavy tails. The exponential power distribution was chosen because it reflects this feature. This choice was also made in \cite{nowak_main}). The tails are heaviest for high values of the log-permeability variance.  In the limit of high variance, the permeability can take a very wide range of values.  Thus, it will often occur that the flow either continues along its main axis, or is diverted.  This behavior is reflected in the statistics by the tails of the distribution.


\section{Conclusion and Discussion}
\label{secV}

The rock heterogeneities exert a significant influence on the flow, from the pore scale up to the kilometer scale.  Calculation of the Darcy pressure statistics depends on an explicit description of the permeability $K$ at the mesoscopic scale. For this work, we have chosen a lognormal distribution.  Simulated annealing can be used to calculate the pressure and flow statistics for any type of realization of the permeability $K$.  An alternative to assuming the lognormal distribution could be the use of multiple-point statistics, a method that directly infers the necessary multivariate distributions from training images \cite{huysmans}, or copulas, which describe the stochastic structure without reference to the corresponding marginal distributions \cite{bardossy}. 

We have shown that our action $S$ (\ref{action}) can be used to apply simulated annealing to calculate Darcy pressure and flow statistics. We have outlined our computational methods in such a way as to make them easily reproducible. Our model was a three-dimensional, bounded domain, with Dirichlet boundary conditions at two ends and no-flow boundary conditions at the remaining four. Our results for the pressure, calculated at two different points in the domain, as well as those for the local and total flow in the main and in a transverse direction al behaved qualitatively as expected. Parametric log-normal and exponential power-law fits were made using Mathematica, all of which passed one-sided Kolmogorov-Smirnov tests at the $95\%$ confidence level. At the moment, simulated annealing is not computationally competitive with the finite-volume method. Its runtime may be improved through the use of the multigrid method, however. Most promisingly, it may be possible to apply the renormalization group as an upscaling technique. Such an application would provide a different take on the problem and may enable the user to run coarse but fast simulations to capture the main characteristics of the pressure and flow statistics.

A major challenge is the extension of the present approach to multiphase flow. For two-phase flow a generalized form of Darcy's law is used:
\begin{equation}\label{Darcy_multiphase}
q_i = -k_{r,i}(S_i)K\nabla p\, ,
\end{equation}
where the subscript $i$ represents the fluid phase (oil or water), $k_{r,i}$ is the relative permeability, and $S_i$ is the pore volume fraction of the fluid phase $i$.  The two volume fractions must sum to one. The total velocity is given by the sum of the individual phase velocities:
\begin{equation}
q = q_o + q_w.
\end{equation}
The rate of change of the saturation $s$ is given by the conservation equation:
\begin{equation}\label{conservation}
\frac{\partial s}{\partial t}=g(s) q \cdot \nabla s\, ,
\end{equation}
where $g(s)$ is a nonlinear function. The constraint (\ref{incompressibility}) still holds for an incompressible fluid. Equation (\ref{Darcy_multiphase}) is similar to Darcy's law for single-phase flow.  An adaptation of the methodology outlined in this paper should be suited to the solution of (\ref{Darcy_multiphase}).  The hyperbolic saturation equation (\ref{conservation}) poses more problems, in particular because its nonlinear nature leads to the formation of a shock in the solution \cite{buckley}. Previous studies \cite{king} indicate how a path integral formulation of this equation can be formed, providing an opportunity for future research.


\begin{acknowledgments} 
MJEW was supported through a Janet Watson scholarship from the Department of Earth Science and Engineering and a studentship in the Centre for Doctoral Training on Theory and Simulation of Materials funded by the EPSRC (EP/L015579/1), both at Imperial College London. 
We acknowledge the support of the Imperial College Research Computing Service \cite{HPC}. 
We thank Dr. Stephan D\"urr for a helpful discussion.
\end{acknowledgments}


\begin{thebibliography}{41}
\bibitem{dimensionAction} Note that the ``cooling parameter'' $T$ does not have the dimensions of a temperature, as the analogy with the cooling of solids suggests.
\end{thebibliography}
\end{document}